# Thermodynamic Model of Hardness: Particular Case of Boron-Rich Solids

**V. A. Mukhanov[a], O. O. Kurakevych[b], and V. L. Solozhenko[a]**

[a] *LPMTM-CNRS, Université Paris Nord, Villetaneuse, France*

[b] *IMPMC, Université P & M Curie, Paris, France*

**Abstract**—A number of successful theoretical models of hardness have been developed recently. A thermodynamic model of hardness, which supposes the intrinsic character of correlation between hardness and thermodynamic properties of solids, allows one to predict hardness of known or even hypothetical solids from the data on Gibbs energy of atomization of the elements, which implicitly determine the energy density per chemical bonding. The only structural data needed is the coordination number of the atoms in a lattice. Using this approach, the hardness of known and hypothetical polymorphs of pure boron and a number of boron-rich solids has been calculated. The thermodynamic interpretation of the bonding energy allows one to predict the hardness as a function of thermodynamic parameters. In particular, the excellent agreement between experimental and calculated values has been observed not only for the room-temperature values of the Vickers hardness of stoichiometric compounds, but also for its temperature and concentration dependencies.

## 1. THERMODYNAMIC MODEL OF HARDNESS

The theory of hardness and design of novel superhard materials are great challenge to materials scientists till now. Diamond-like and boron-rich compounds of light elements (Fig. 1a) take a particular place in this research, since the hardest known phases have mainly these two structural types [1].

Many attempts have been made to predict hardness using the structural data and such characteristics as bulk ($B$) and shear ($G$) moduli, specific bond energy, band gap ($E_g$), density of valence electrons (i.e. the number of valence electrons per unit volume), etc. [2–7]. Up to date the best correspondence between the calculated and experimental values of hardness has been achieved in the recent works [3, 4, 8–10]. In all cases, the final accuracy is about 10% for hard and superhard phases, i.e. at the level of experimental errors.

According to our model [8–10], the hardness of a phase with isodesmic structure[1] is proportional to the atomization energy, which may be considered as a characteristic of the bond rigidity (for clarity, we will use the standard values of Gibbs energy of atomization $\Delta G^{o}_{at}$ ), and is in inverse proportion to the molar volume of a phase and to the maximal coordination number of the atoms. The value defined in such a way has the dimensions of pressure. The plasticity of materials can be taken into account by empirical coefficient $\alpha$. In general case the polarity of bonds leads to the hardness decrease, which may be clearly seen in the sequence of isoelectronic analogues of diamond, i.e. diamond (100 or 115 GPa) [11, 12]—cubic boron nitride cBN (62 GPa) [13]—BeO (13 GPa) [3,14]—LiF (1.5 GPa) [3, 14]. This factor has been evaluated by empirical coefficient $\beta$, which is the measure of the bond covalence.

The equation that allows calculating the Vickers hardness ($H_V$) of crystals at 298 K is

$$H_V = \frac{2\Delta G^{o}_{at}}{VN}\alpha\beta\varepsilon, \qquad (1)$$

where $V$ is the molar (atomic) volume (cm$^3$ mole$^{-1}$); $N$ is the maximal coordination number[2]; $\alpha$ is the coefficient of relative (as compared to diamond) plasticity; $\beta$ is the coefficient corresponding to the bond polarity (see below); $\varepsilon$ is the ratio between the mean number of valence electrons per atom and the number of bonds with neighboring atoms ($N$)[3]; $\Delta G^{o}_{at}$ is the standard Gibbs energy of atomization (kJ mole$^{-1}$) of compound $X_mY_n$:

$$\Delta G^{o}_{at\,X_mY_m} = m\Delta G^{o}_{at\,X} + n\Delta G^{o}_{at\,Y} - \Delta G^{o}_{f\,X_mY_m}, \tag{2}$$

where $\Delta G^{o}_{f\,X_mY_m}$ is the standard Gibbs energy of formation of $X_mY_n$; $\Delta G^{o}_{at\,X}$ and $\Delta G^{o}_{at\,Y}$ is the standard Gibbs energies of atomization of elements $X$ and $Y$.

---

[1] Isodesmic structure is characterized by the similar bond strength in all direction.

[2] For some compounds of very complex structure, such as boron-rich solids, we will use a mean/effective value.

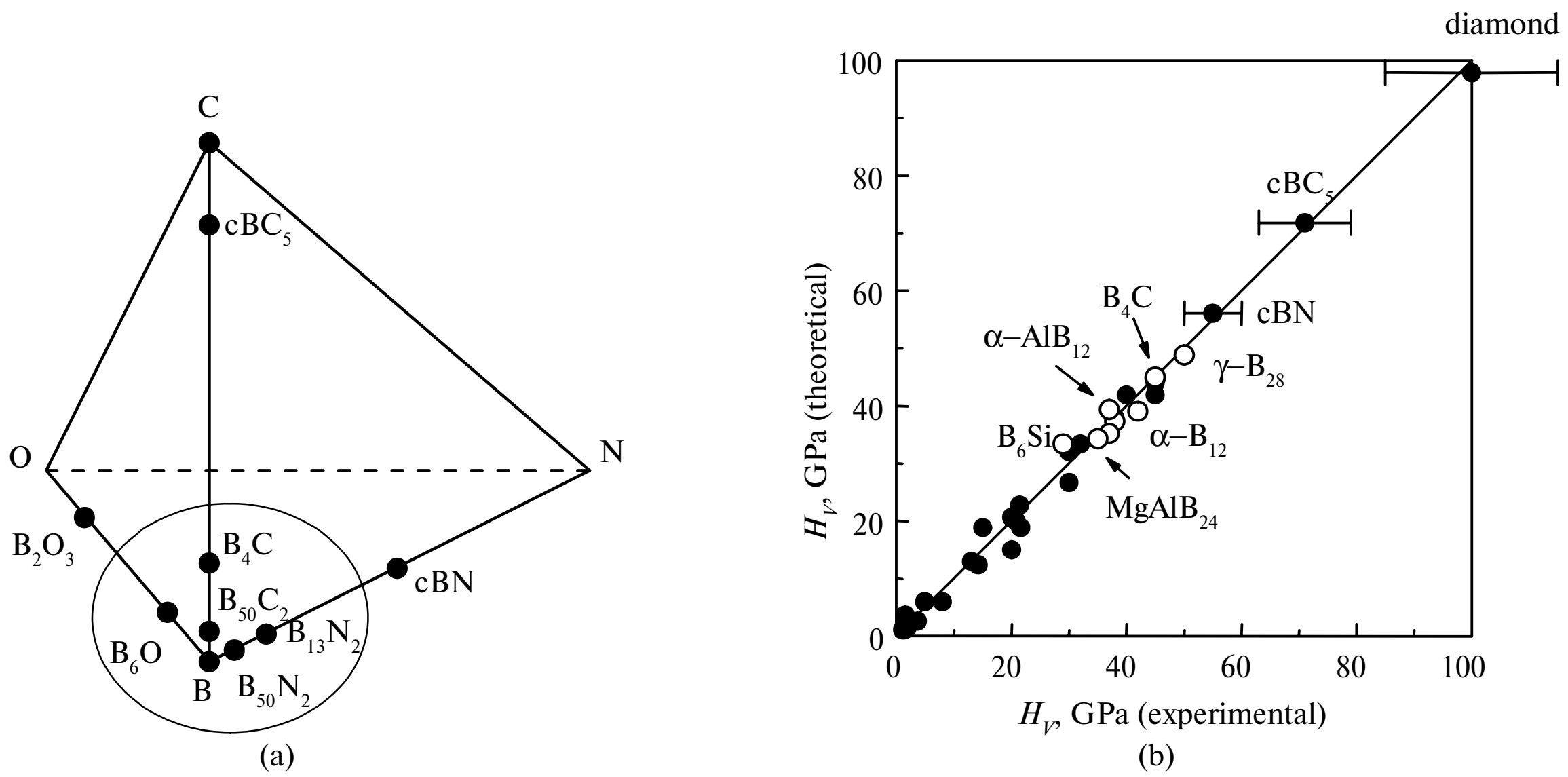

**Fig. 1.** Principal hard and superhard phases on the B–C–N–O concentration tetrahedron. The boron-rich phases are surrounded by an oval line (a). A comparison of experimental values of Vickers hardness of various phases with corresponding values calculated as a function of Gibbs energy of atomization in the framework of the thermodynamic model of hardness (Eq. (1)). The open circles correspond to the boron-rich solids, while all other compounds/phases are presented by solid circles (b).

Coefficient $\alpha$ has been estimated from the experimental values of $H_V$ for diamond, dSi, dGe and dSn. For the elementary substances and compounds of second period elements $\alpha$ equals 1, while for other periods ($\geq 3$) $\alpha$ makes 0.7. This coefficient reflects the decrease of the bond strength [4] for the elements of periods $\geq 3$. The precise estimation of this coefficient, reflecting the presence of large inner electron core and multiple non-occupied $d$- and $f$-orbitals, is outside of the purposes of this study.

Coefficient $\beta$ (square of the covalence $f$) has been calculated by the equation

$$\beta = \left(\frac{2\chi_Y}{\chi_Y + \chi_X}\right)^2; \tag{3}$$

where $\chi_X$, $\chi_Y$ are the electronegativities of the elements by Pauling, $\chi_X > \chi_Y$ [15]. For elementary substances $\beta = 1$. In fact, even the presence of small amounts of foreign atoms in the structure should cause the remarkable decrease of hardness, as it can be seen by the example of boron-doped diamonds [16] (from 90–110 GPa for pure single-crystal diamond down to 70–80 GPa for single crystals of boron-doped diamond; i.e. down to ~ 75–85% of the initial value, which well match the square of bond ionicity $\beta = 0.79$ for B–C bonds).

For the refractory crystalline compounds the values of hardness calculated by Eq. (1) are in a very good agreement (in the most cases less than 4 GPa of discrepancy, i.e. < 7%) with the experimental values [1, 3, 4, 11–14, 17–21] (Fig. 1b[4]).

[3] The use of this coefficient allows one to establish the hardness of the $A^{I}B^{VII}$ ($\varepsilon = 1/N$) and $A^{II}B^{VI}$ ($\varepsilon = 2/N$) compounds, i.e. LiF, NaCl, BeO, ZnS, MgO, etc.

[4] The considered compounds/phases are diamond, Si, Ge, dSn, SiC, cBN, wBN, c-$BC_2N$, α-rh B, β-rh B, $B_4C$, $B_6O$, TiC, $Si_3N_4$, BeO, TiN, $Al_2O_3$, quartz, coesite, stishovite, WC, $ReB_2$, LiF, $Al_2SiO_4F_2$, $KAlSi_3O_8$, $Ca_5(PO_4)_3F$, $CaF_2$, $CaCO_3$, BAs, BP, AlN, AlP, AlAs, AlSb, GaN, GaP, GaAs, GaSb, InN, InP, InAs, InSb, ZnS, ZnSe, ZnTe, ZnO.

Using Eq. (1) it is possible to calculate the hardness of dense phases with three-dimensional structures that have not been synthesized to present time, e. g., $C_3N_4$ with the $Si_3N_4$ structure [5], $CO_2$ with the $\alpha$-$SiO_2$ structure, hp-$B_2O_3$ with the $Al_2O_3$ structure [17], and a number of diamond-like phases of the B–C system [22]. The advantage of the proposed method is that only the maximum coordination number is used as a structural data [8]. In this case the molar volumes may be calculated from the covalent radii of the elements, while $\Delta G_f^o$ values (usually the negligible term as compared with $\Delta G_{at}^o$ of the elements) of the phases may be fixed to the standard Gibbs energies of the formation of known compounds in the corresponding binary systems, i.e. $C_2N_2$, $CO_2$, $\beta$-$B_2O_3$, $B_4C$, respectively [9]. The applicability of this method for estimating the hardness of hypothetical compounds has been recently illustrated by the example of diamond-like $BC_5$ (c$BC_5$), a novel superhard phase synthesized under high pressures and temperatures [18, 23]. Vickers hardness of this phase has been calculated to be 70.6 GPa (table), which is in excellent agreement with the experimental value $H_V = 71$ GPa.

The theoretical (Eq. (1)) and experimental values of Vickers hardness $H_V$. The starting data (free energy of atomization, density, coordination number and electronegativity) as well as some intermediate values of calculation are also given

| Phase* | $\Delta G_f$**, kJ/mole-at | $\Delta G_{at}$ El [15] | $\Delta G_{at}$, kJ/mole-at | ρ, g/cm$^3$ | $V$, cm$^3$/mole-at | $N$ | $2\Delta G_{at}/NV$ GPa | Electronegativity [15] | | | α | β | $H_V$ | |
|---|---|---|---|---|---|---|---|---|---|---|---|---|---|---|
| | | | | | | | | El | anion | cation | | | theor. | exp. |
| | | | | | | | Boron modifications | | | | | | | |
| $\alpha$-$B_{12}$ | 0 | 0 | 518.8 | 2.447 | 4.418 | 5 | 47.0 | | 2.04 | 2.04 | 1 | 1 | 47.0 | 42 [39] |
| ″ | – | – | – | – | – | 6 | 39.1 | | – | – | – | – | 39.1 | |
| $\beta$-$B_{106}$ | 0 | 0 | 518.8 | 2.334 | 4.632 | 5 | 44.8 | | 2.04 | 2.04 | 1 | 1 | 44.8 | 45 [40] |
| T-$B_{192}$ | 0 | 0 | 518.8 | 2.340 | 4.620 | 5 | 44.9 | | 2.04 | 2.04 | 1 | 1 | 44.9 | |
| $\gamma$-$B_{28}$ [42] | 0 | 0 | 518.8 | 2.544 | 4.249 | 5 | 48.8 | | 2.04 | 2.04 | 1 | 1 | 48.8 | 50 [58] |
| dB | 0 | 0 | 518.8 | 2.548 | 4.243 [8] | 4 | 61.1 | | 2.04 | 2.04 | 1 | 1 | 61.1 | |
| ″ | – | – | – | 2.178 | 4.963 [18] | – | 52.3 | | | – | – | – | 52.3 | |
| $\alpha$-Ga type | 0 | 0 | 518.8 | 2.810 | 3.847 [42] | 7 | 38.5 | | 2.04 | 2.04 | 1 | 1 | 38.5 | |
| ″ | – | – | – | – | – | 6 | 45.0 | | 2.04 | 2.04 | 1 | 1 | 45.0 | |
| | | | | | | | Compounds of the diamond structural type | | | | | | | |
| cBN | –120.15 | 455.563 | 607.3 | 3.489 | 3.555 | 4 | 85.4 | 3.04 | 3.04 | 2.04 | 1 | 0.645 | 55.1 | 62 [13] |
| c$BC_5$ [18] | 0 | 671.26 | 645.8 | 3.267 | 3.612 | 4 | 89.4 | 2.55 | 2.55 | 2.04 | 1 | 0.79 | 70.6 | 71 |
| BP | –47.4 | 278.3 | 446.0 | 2.970 | 7.034 | 4 | 31.7 | 2.19 | 2.19 | 2.04 | 1 | 0.93 | 29.5 | 33 [59] |
| | | | | | | | Compounds of the $\alpha$-$B_{12}$ structural type | | | | | | | |
| $B_6O$ | –93 | 231.7 | 572.6 | 2.575 | 4.474 [60] | 5 | 51.2 | 3.44 | 2.74 | 2.04 | 1 | 0.729 | 37.3 | 38 [21] |
| ″ | 0 | – | 479.6 | – | – | – | 42.9 | – | – | – | – | – | 31.2 | |
| $B_4C$ | –12 | 671.3 | 561.3 | 2.507 | 4.407 | 5 | 50.9 | 2.55 | 2.30 | 2.04 | 1 | 0.884 | 45.0 | 45 [46] |
| ″ | 0 | – | 549.3 | – | – | – | 49.9 | – | – | – | – | – | 44.1 | |
| $B_9C$ | 0 | 671.3 | 534.0 | 2.282 | 4.789 | 5 | 44.6 | 2.55 | 2.30 | 2.04 | 1 | 0.88 | 39.4 | |
| $B_{13}N_2$ [48] | –20 | 455.6 | 530.4 | 2.666 | 4.214 [47] | 5 | 50.3 | 3.04 | 2.54 | 2.04 | 1 | 0.794 | 39.9 | |
| $B_4Si$ | 0 | 411.3 | 497.3 | 2.425 | 5.882 | 5 | 33.8 | 1.90 | 2.04 | 1.97 | 1 | 0.97 | 32.6 | 27 [61] |
| $B_6P$ | 0 | 278.3 | 484.4 | 2.583 | 5.300 | 5 | 36.6 | 2.19 | 2.12 | 2.04 | 1 | 0.962 | 35.2 | 37 |
| $B_6As$ | 0 | 261 | 482.0 | 3.570 | 5.593 | 5 | 34.5 | 2.18 | 2.12 | 2.04 | 1 | 0.962 | 33.2 | |

(Contd.)

| | | | | | | | | | | | | | | |
|---|---|---|---|---|---|---|---|---|---|---|---|---|---|---|
| Compounds of the $\beta$-$B_{106}$ structural type | | | | | | | | | | | | | | |
| $B_{25}Mg_2$ [62] | 0 | 127 | 488.7 | 2.488 | 4.747 | 5 | 41.2 | 1.31 | 2.04 | 1.68 | 1 | 0.81 | 33.5 | |
| $B_{19.7}Mg$ [63] | 0 | 127 | 499.9 | 2.416 | 4.744 | 5 | 42.1 | 1.31 | 2.04 | 1.68 | 1 | 0.81 | 34.2 | |
| $AlB_{31}$ [64] | 0 | 285.7 | 511.5 | 2.411 | 4.693 | 5 | 43.6 | 1.61 | 2.04 | 1.83 | 1 | 0.89 | 39.0 | |
| $B_{36}Si$ [65] | 0 | 411.3 | 515.9 | 2.343 | 4.813 | 5 | 42.9 | 1.90 | 2.04 | 1.97 | 1 | 0.97 | 41.4 | |
| Compounds of the T-$B_{52}$ structural type | | | | | | | | | | | | | | |
| $B_{50}C_2$ | 0 | 671.3 | 524.7 | 2.395 | 4.533 | 5 | 46.3 | 2.55 | 2.30 | 2.04 | 1 | 0.884 | 40.9 | |
| $B_{50}N_2$ | 0 | 455.6 | 516.4 | 2.454 | 4.455 | 5 | 46.4 | 3.04 | 2.50 | 2.04 | 1 | 0.808 | 37.4 | |
| $B_{50}B_2$ | 0 | 0 | 518.8 | 2.383 | 4.536 [48] | 5 | 45.7 | | 2.04 | 2.04 | 1 | 1 | 45.7 | |
| Compounds of the T-$B_{192}$ structural type | | | | | | | | | | | | | | |
| $\alpha$-$AlB_{12}$ | 0 | 285.7 | 500.9 | 2.650 | 4.549 | 5 | 44.0 | 1.61 | 2.04 | 1.83 | 1 | 0.894 | 39.4 | 37 |
| Other boron-rich compounds | | | | | | | | | | | | | | |
| $\gamma$-$AlB_{12}$ | 0 | 285.7 | 500.9 | 2.560 | 4.709 | 5 | 42.5 | 1.61 | 2.04 | 1.83 | 1 | 0.894 | 38.1 | |
| $oB_6Si$ [66] | 0 | 411.3 | 502.5 | 2.399 | 5.598 | 5 | 35.9 | 1.90 | 2.04 | 1.97 | 1 | 0.97 | 34.7 | 29 [67] |
| $MgAlB_{14}$ | 0 | 199.4 | 478.9 | 2.660 | 4.761 | 5 | 40.2 | 1.46 | 2.04 | 1.75 | 1 | 0.85 | 34.3 | 35 [68] |
| $WB_4$ | 0 | 807.1 | 576.5 | 10.193 | 4.456 | 5 | 51.7 | 2.36 | 2.20 | 2.04 | 1 | 0.93 | 47.9 | 46 [69] |

Notes: * Hypothetical phases are given in italics.

** For the most of compounds the estimation of $\Delta G_f$ is given using the thermodynamic data of known phases. For the boron-rich compounds with high boron content (> 85 at % of B) it has been fixed to 0.

Equation (1) also allows one to calculate the values of hardness at various temperatures by introducing the linear approximation of the temperature dependence of $\Delta G_{at}(T)$, i.e.

$$\Delta G_{at}(T) = \Delta G_{at}(300)\cdot[1 - (T - 300)/(T_{at} - 300)], \tag{4}$$

where $T_{at}$ is the temperature of atomization[5]; as well as by introducing the temperature dependences of molar volumes $V(T)$. The theoretical simulation [8—10] shows a good agreement with the experimental data on the temperature dependences of Vickers and/or Knoop hardness for diamond, cBN, $B_4C$, $ReB_2$ and $Al_2O_3$ (Figs. 2a, 2b) in comparison with experimental data. The theoretical values of hardness have been calculated by equation

$$H(T) = H(300)\frac{\Delta G_{at}(T)V(300)}{\Delta G_{at}(300)V(T)}. \tag{5}$$

At relatively high temperatures (~ 0.3—0.5 $T_{at}$) this equation gives 10—15 % higher values than the observed ones (bold lines in Figs. 2a, 2b), that should be attributed to the increase of materials' plasticity due to the intensification of the surface and bulk diffusion [24]. The influence of the temperature on plasticity (coefficient $\alpha$) can be taken into account by the following empirical equation:

$$\alpha(T) = \alpha(300)\left(1 - \frac{k(T)}{k(\infty)}\right) = \alpha(300)\left(1 - e^{-\frac{2/3T_{melt}}{T}}\right), \tag{6}$$

that supposes the Arrhenius-type temperature dependence of the dislocation propagation constant $k(T)$ (following Ref. [25], the activation energy was set to $2/3RT_{melt}$). This term allows a decrease in the discrepancy between experimental and calculated data down to the level of experimental error (dashed lines in Figs. 2a, 2b). Our model describes the lowest possible decrease of hardness in the case of a material of a fixed microstructure. This explains the excellent agreement between the experimental and theoretical $H_V(T)$ curves for single crystals (Fig. 2a); while in the case of polycrystalline ceramic materials some deviations (often non-monotone) may occur due to the temperature-induced microstructure changes.

[5] For diamond and cBN the corresponding temperatures of sublimation are 4300 K and 3300 K, respectively.

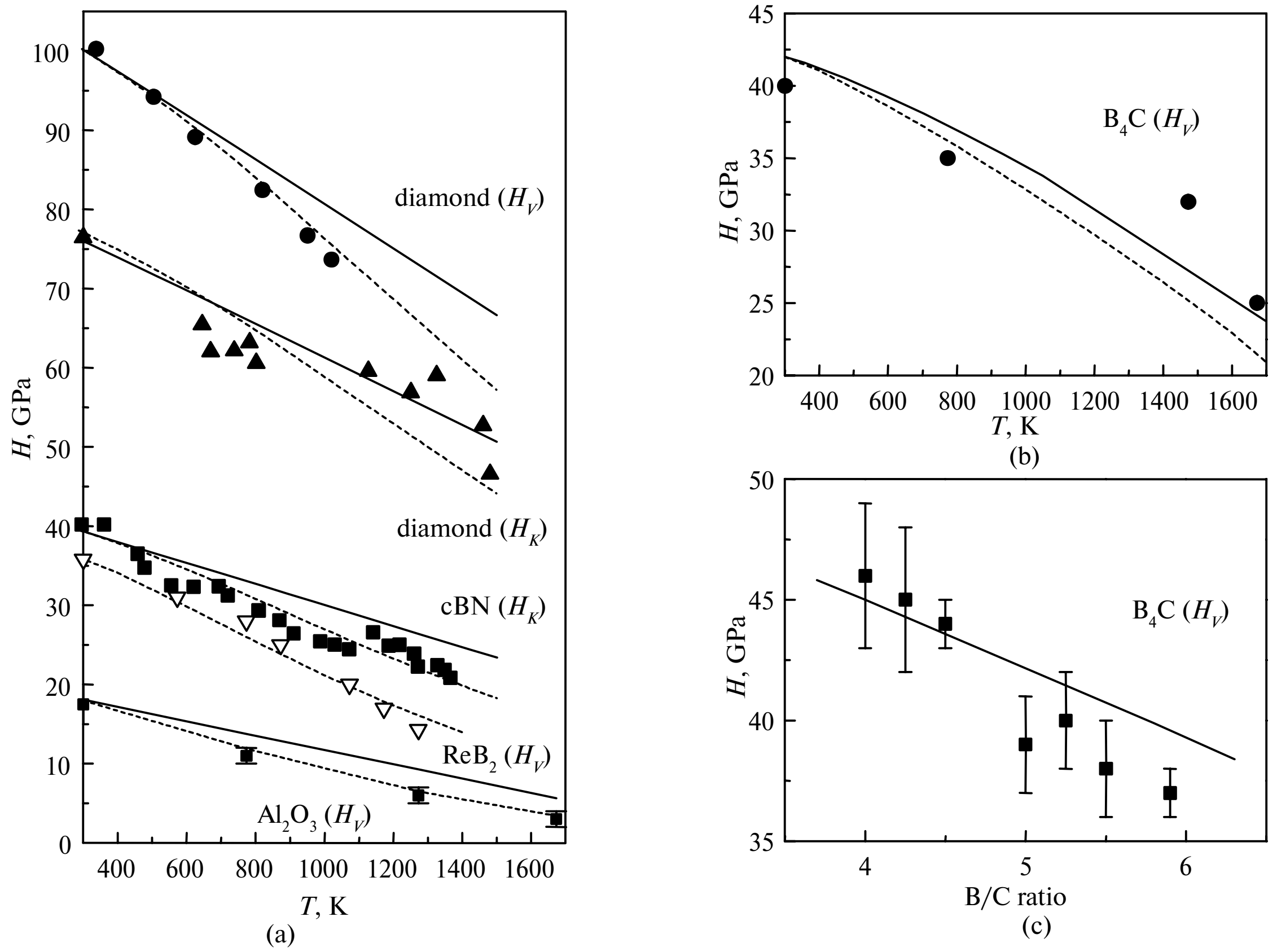

**Fig. 2**. Temperature dependence of hardness of single-crystal diamond, polycrystalline cBN (mean particle size of 5 μm), single-crystal $ReB_2$, and $B_4C$-, SiC- and $Al_2O_3$-based ceramics (a, b). The symbols represent the experimental data obtained by static indentation [52–57]. The lines show the results of calculation using Eq. (5) under assumption that $\alpha$ = const (solid line) and using Eq. (6) for $\alpha$ (dashed line). The concentration dependence of boron carbide hardness (c). The symbols represent experimental data [51], while solid line shows the results of calculation using Eq. (1). The crystallographic density of corresponding carbides has been evaluated using the lattice parameter data reported in [51].

Our model has justified the previous suggestions about the increase of hardness with pressure [26]. Because of the lack of reliable data on $\Delta G_{at}$ at very high pressures, the prediction cannot be easily made using equations (1) (the *ab initio* calculations of corresponding thermodynamic parameters could be useful in this case). However, according to [10] where the non-monotone correlation between $H_V$ and bulk modulus $B$ has been explained in the framework of the same thermodynamic model under the assumption of the similar nature of energy stocked by chemical bonds during indentation and compression, the pressure dependence of hardness is the same (up to a constant depending on material only) as the pressure dependence of $B$, i.e.

$$H_V(p) = \mathrm{const}B(p). \tag{7}$$

Equation (7) allows one to suggest that the hard phases even with relatively low bulk moduli may show a remarkable hardness increase with pressure. Each material is expected to increase its hardness when the pressure is applied, however, the phase transformations accompanied by an increase of coordination number could prevent an infinite increase and cause a drop of hardness at a transformation pressure.

Some compounds with relatively high hardness at ambient pressure and relatively high pressure derivative of bulk modulus (as compared to diamond) under pressure become harder more rapidly than diamond (as soon as the structural phase transformations increasing the coordination number occurs); that allows some of them (e. g., $B_6O$) to reach the diamond hardness at very high pressures [26]. It is interesting to note that graphite, a very soft material at ambient conditions, may reach the diamond hardness at the lower pressure than many other materials. This fact is in excellent agreement with the experimental and theoretical results reported in [27, 28] on the formation of "superhard graphite" that can scratch a single-crystal diamond, and allows us to suggest that other ordered [29–31] and disordered [32–35] graphite-like phases should show similar behavior under high pressure, even if the "compressed state" is not always recoverable at ambient pressure.

## 2. HARDNESS OF BORON-RICH SOLIDS

One more advantage of the proposed method is the possibility to easily estimate the hardness of various forms of boron and its compounds ($B_4C$, $B_6O$, $B_{13}N_2$, etc., see Fig. 3a, table), which is rather complicated by using other methods because of extreme complexity of boron-related structures and a large number of atoms in a unit cell. Usually the "non-ionic" contribution to hardness ($2\Delta G_{at}/NV$) is close to that of pure boron. However, the role of ionicity is not so clear because of the strong delocalization of chemical bonds. Thus, in our calculations for boron-rich compounds we have taken the mean value of electronegativities of all atoms connected to $B_{12}$ icosahedron as an $\chi$ value for anion (or cation). For almost all boron-rich compounds the mean coordination number has been fixed to $\langle N \rangle = 5$[6]; and only for $\alpha$-$B_{12}$ to 6 because the half of its icosahedral B-atoms have coordination number 7 due to the formation of three-center electron-deficient bondings.

Here we should also remark that the microstructure developed during various synthesis procedures [36] and even the influence of the single-crystal purity may significantly affect the hardness [37]. Thus, the poor/lacking data on the hardness of some boron-rich compounds may cause the significant under- or overestimation of experimental $H_V$-values.

### *2.1. Boron Polymorphs*

Boron is known to be the hardest element next to carbon [1, 38]. The experimental values of hardness for $\alpha$-$B_{12}$ ($H_V = 42$ GPa [39]) and $\beta$-$B_{106}$ ($H_V = 45$ GPa as the maximal hardness for samples remelted at ambient pressure [40, 41] and 44(4) GPa for samples remelted at ~ 5 GPa, according to our unpublished results) phases are in good agreement with the values (39.2 and 43.8 GPa, respectively) calculated in the framework of the thermodynamic model of hardness. The hardness of recently synthesized superhard high-pressure boron phase, orthorhombic $\gamma$-$B_{28}$ [42, 43], was found to be 50 GPa [19], which also agrees well with the calculated value of 48.8 GPa. Our model suggests that $\gamma$-$B_{28}$ has the highest hardness among the known crystalline modifications of boron (as well as the lowest compressibility [44]) because of its highest density. The hardest polymorph is expected to be hypothetical diamond-like boron, a strongly metastable covalent phase, which, probably, could be stabilized (e.g., by quenching down to low temperatures) if the activation barrier of its transformation into conventional boron phases is high enough. Using different estimations of atomic volume (table), the expected hardness of dB should vary between 52 and 61 GPa. The hardness of tetragonal polymorph T-$B_{192}$ [45] has not been ever reported. However, our calculations have shown that it should be the same as that of rhombohedral $\beta$-$B_{106}$.

### *2.2. Boron-rich Solids of the $\alpha$-$B_{12}$ Structural Type*

Although the $\alpha$-$B_{12}$ phase is metastable at ambient pressure [42], the small amount of non-metal contaminations (C, O, N, Si, etc.) stabilizes the boron-rich compounds of the $\alpha$-$B_{12}$ structural type. The calculated values of Vickers hardness for $B_4C$ and $B_6O$ are 44 and 38 GPa, respectively; that is in a very good agreement with the experimental data for single crystal $B_4C$ ($H_V = 45$ GPa [46]) and polycrystalline $B_6O$ ($H_V = 38$ GPa [21]). The lower value of hardness for $B_6O$ as compared to $B_4C$ may be explained by the higher ionicity of B–O bonds than of B–C bonds. The estimation of hardness for the recently synthesized rhombohedral boron subnitride $B_{13}N_2$ [47–49] has given $H_V = 40.3$ GPa[7] that allows ascribing $B_{13}N_2$ to superhard phases. Its relatively high bulk modulus comparable to those of $B_4C$ and $B_6O$ additionally confirms this suggestion [50].

Boron carbide $B_4C$, a very hard substance, which may be produced at ambient pressure, is, in fact, a kind of a solid solution of carbon in boron, i.e. $B_{4+x}C_{1-x}$, having a wide concentration range of stability. Using our model of hardness, we have also succeeded to calculate the concentration dependence for the Vickers hardness of $B_{4+x}C_{1-x}$ (Fig. 2c), which is in a satisfactory agreement with experimental data reported in [51].

---

[6] Since it is difficult to decide whether the B-atoms or $B_{12}$ icosahedra should be considered as structural units, the approximate mean values of coordination number has been taken, which give the best agreement between calculated and experimental data.
[7] The $2\Delta G^{o}_{at}/NV$ value has been set to a mean (~ 51 GPa) of corresponding values for $B_6O$ and $B_4C$; $\beta = 0.79$.

The calculated hardness of the $\alpha$-$B_{12}$-type compounds with the elements of the 3rd and higher periods somewhat decreases due to the high concentration of the polar (partly ionic) bonds. At the same time, most of the phases not only belong to the hardness range assigned to the "hard phases", but also are close to its upper limit (Fig. 3a).

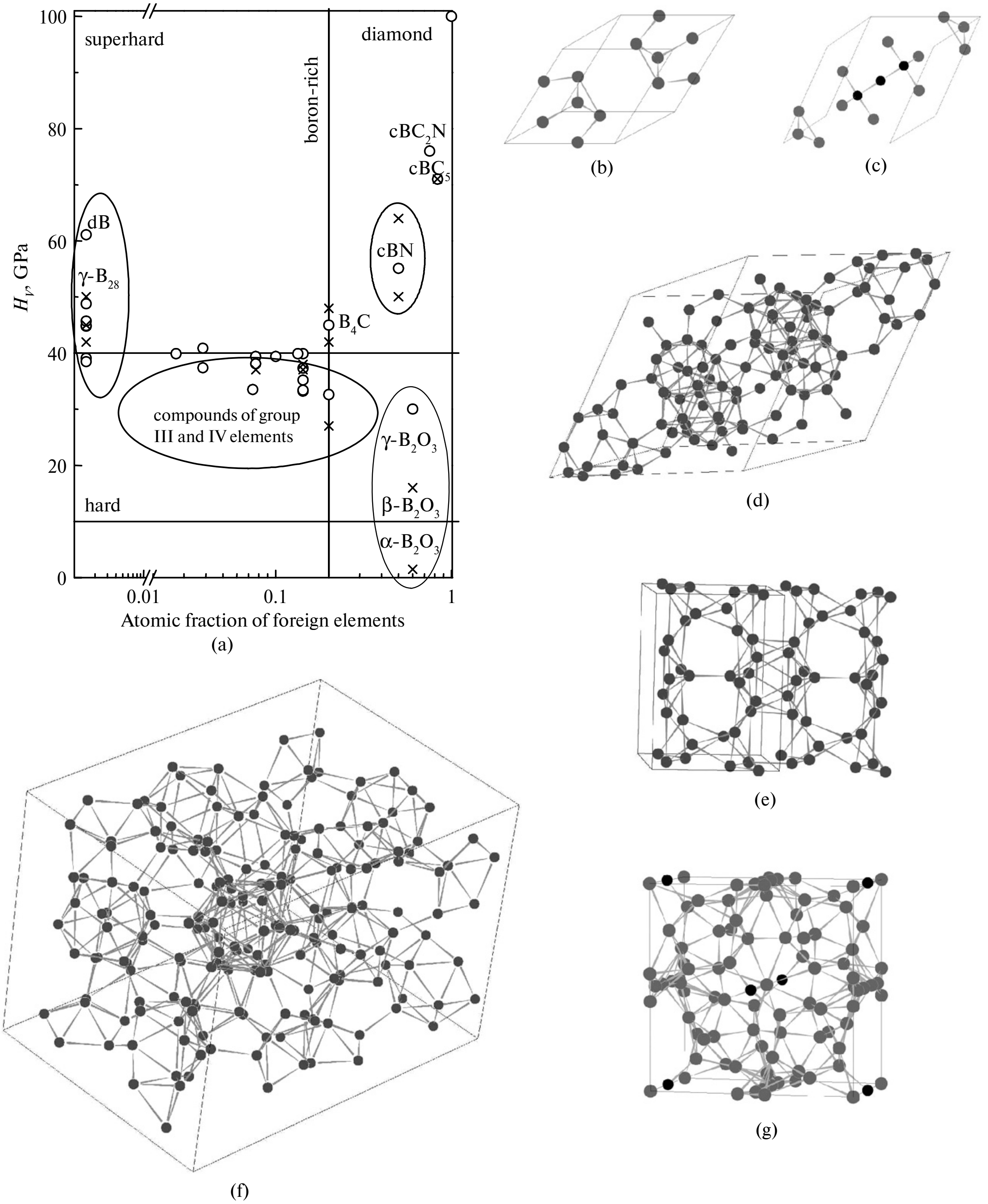

**Fig. 3**. Theoretical (○) and experimental (×) values of Vickers hardness of boron-containing compounds as a function of the atomic fraction of foreign element(s) (a). The approximate boundaries of hard ($H_V > 10$ GPa), superhard ($H_V > 40$ GPa), and boron-rich (boron content  80 at %) phases are given by horizontal and vertical lines. Principal structural types of boron-rich compounds related to known and hypothetical modifications of boron (b–g); i.e. rhombohedral $\alpha$-$B_{12}$ (b, c) and $\beta$-$B_{106}$ (d); orthorhombic $\gamma$-$B_{28}$ (e); tetragonal T-$B_{192}$ (f), and T-$B_{52}$ (g).

### 2.3. *Boron-rich Solids of the β-$B_{106}$ Structural Type*

The β-$B_{106}$ phase is the only thermodynamically stable phase of boron at pressures up to few GPa's. Very small amounts of foreign elements, especially metals, give a rise to a number of boron-rich compounds of the β-$B_{106}$ structural type (table). Their hardness is expected to be lower just because of the partial ionicity of the chemical bonds.

### 2.4. *Boron-rich Solids of the T-$B_{52}$ Structural Type*

The hypothetical T-$B_{52}$ phase may be stabilized only by a small amount of nitrogen or carbon atoms as compounds $B_{50}N_2$ and $B_{50}C_2$. The hardness of these phases have not been experimentally studied to the present day, while our calculations show that they should have hardness between $B_6O$ and $B_4C$, two common superhard phases of the α-$B_{12}$ type. The recent studies of the B–BN system under high pressure [48] have revealed the stabilization of the phase that, most probably, is a solid solution $B_{50}N_{2-x}B_x$, with $x \approx 2$. Its hardness is supposed to be very close to that predicted for T-$B_{52}$, i.e. 46 GPa.

## 3. CONCLUSIONS

Thus, it has been found that the hardness of solids is directly related to their thermodynamic and structural properties. The formulated equations may be used for a large number of compounds with various types of chemical bonding and structures. The proposed method allows estimating the hardness and compressibility of various hypothetical compounds using the data on the Gibbs energy of atomization of elements and covalent/ionic radii. The applicability of the approach to the prediction of hardness has been illustrated by examples of the recently synthesized superhard diamond-like $BC_5$ and orthorhombic modification of boron, γ-$B_{28}$. In the framework of the proposed method we have calculated the hardness of a large number of boron-rich solids and found that it strongly depends on the electronegativity of atoms incorporated into boron lattice.

The authors are grateful to the Agence Nationale de la Recherche for the financial support (grant ANR-05-BLAN-0141).